# The $A_T$ emission of KCl:Tl interpreted as a double $A_T+A_X$ emission

D. Mugnai, G. P. Pazzi, P. Fabeni, and A. Ranfagni

*"Nello Carrara" Institute of Applied Physics,*
*CNR Florence Research Area, Via Madonna del Piano 10, 50019 Sesto Fiorentino,*
*Florence, Italy*

In this paper we analyze the emission spectra of $Tl^+$-like impurity centers in alkali-halide crystals, in relation to their anomalous behavior in the decay times. The anomaly, which has been experimentally evidenced since 1992 for $Pb^{2+}$ impurity [1], consists of a gradual change in the decay time, from the faster component (at ns) to the slower (at ms) one, by assuming all the intermediate values.

In spite of the existence of some theoretical models, which have been performed in order to explain this anomaly [2-6], the problem appears to be still an open question.

We present here a model in which we consider the possibility that the $A_T$ emission of KCl:Tl can be considered as a double emission, namely a superposition of the $A_T$ and $A_X$. As we shall see further on, in our model a crucial role is played by the tunneling process.

In the case of $Tl^+$ impurity the main characteristic of the system is represented by the co-existence of two kinds of Jahn-Teller minima, on the adiabatic potential energy surface (APES) relative to the $^3T_{1u}^*$ excited state of the impurity from which the $A_T$ and $A_X$ emissions originate. This coexistence is allowed by the spin-orbit mixing between the triplet $^3T_{1u}$ and the singlet $^1T_{1u}$ from which the $^3T_{1u}^*$ and $^1T_{1u}^*$ states originate.

By considering only the tetragonal $Q_2$ and $Q_3$ coordinates, we can visualize the situation by means of the potential map of the $^3T_{1u,z}^*$ state [5,6] shown in Fig. 1, where the green lines represent the reaction paths for non-radiative transitions from $T_z^*$ to $X$ minima. In Fig. 1 the potential profile along one reaction path is also sketched. As can be seen by looking at the inset in Fig. 1, in order to determine the decay time of the emission, a fundamental role is played by the potential barrier between $T^*$ and $X$ minima, the height of which is $E_T$. At a very low temperature the barrier can be crossed by tunneling, while the crossing can take place for thermal overcoming at a relatively high temperature.

Under the assumption that, after optical excitation, the system relaxes preferentially in the $T^*$ minima, the criticality in the delay time (to which we have to add the lifetime in the state) determines the relaxation time of the $A_X$ emission originating by optical transition from $X$ minima. This is in qualitative agreement with the observation reported in [7], according to which the 300 nm emission of KCl:Tl at low temperatures (up to 80 K) presents both a fast component (of the order of $10^{-8}$ s), and a slow component (of the order of $10^{-3}$ s). The fast component is presumably emitted mainly from the $T^*$ minima, while the slow component is emitted from $X$ minima after overcoming the barrier.

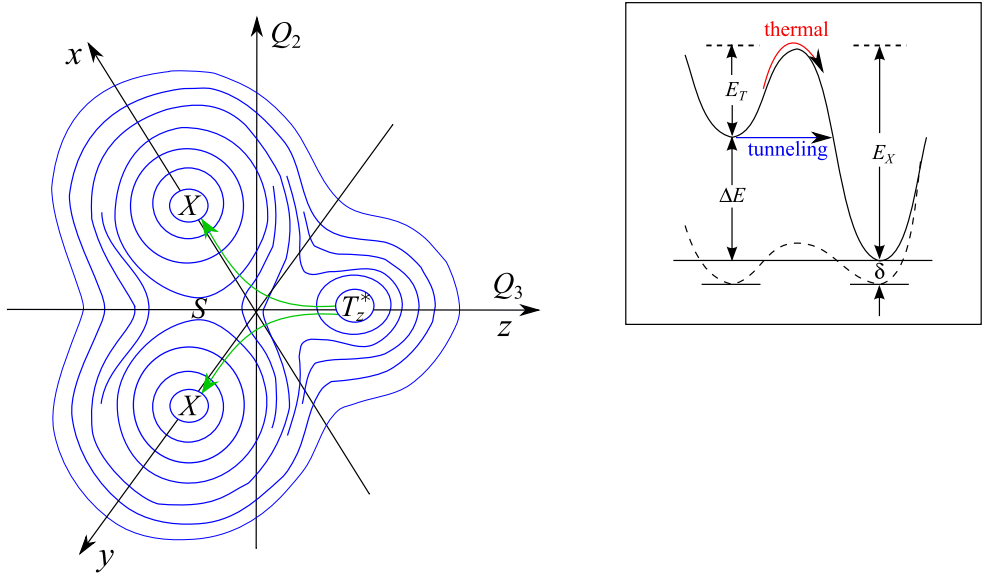

**Fig.1**. Map of the $^3T^*_{1u,z}$ potential surface, in the $Q_2$-$Q_3$ subspace of tetragonal coordinate, showing the coexistence of two equivalent $X$ minima (corresponding to the cluster elongations along the $x$ and $y$ axes, while $S$ is a saddle point between them) with a higher-lying $T^*_z$ minimum elongated along the $z$-axis. The green lines represent paths for non-radiative transitions from $T^*_z$ to $X$ minima. In the inset is shown a sketch of the potential profile of the $^3T^*_{1u,z}$ state, with the underlying $^3A_{1u}$ trap level, as it appears along one of the non-radiative transition-path. The arrows in the barrier region represents crossing by tunneling or thermal overcoming.

A confirmation of this model can be found in the results relative to the degree of linear polarization [8], which is very evident in the fast component, while it is less evident in the slow component, which tends to disappear or becomes negative. For exciting light polarized along $z$, the loss of polarization can be attributed to non-radiative transitions from $T^*_z$ minima, which are elongated along the $z$-axis, and to $X$ minima elongated along the $x$- and $y$-axis, respectively (see Fig. 1)

In order to perform a quantitative test of the model, it is sufficient to consider the cross section of the $^3T^*_{1u,z}$ APES along the $Q_3$ coordinate since, in this way, we can obtain a useful description of the potential in the region of the tetragonal minimum $T^*_z$ and of the barrier.

By denoting the spin-orbit coupling constant with $\zeta$, and by putting $y=E/\zeta$, $x \equiv x_3 = (b/2\sqrt{3}\,\zeta)\,Q_3$, the cross section is given by [5]

$$y = -x - \frac{1}{4} + g - \left[\left(3x - \frac{1}{4} - g\right)^2 + \frac{1}{2}\right]^{\frac{1}{2}} + Ax^2 \qquad (1)$$

where $g=G/\zeta$ is the exchange integral normalized to $\zeta$, $A=12\,(1-\beta)\zeta/b^2$, $\beta$ is a quadratic term, and $b$ is the electron-lattice coupling constant for the tetragonal modes. The model described by Eq. (1) results to be dependent only on the two dimensionless parameters $A$ and $g$. In Fig. 2 some cross sections of $y$ computed by Eq. (1) are shown, for some values of $g$ comprised within the 0-0.4 range, and for $A=3.5$. In the same figure, one height of the barrier, $E_T/\zeta$, is also indicated (for $g=0.1$), together with the separation $(\Delta E + \delta)/\zeta$ of the $T_z^*$ minimum over the trap level at $g=0$.

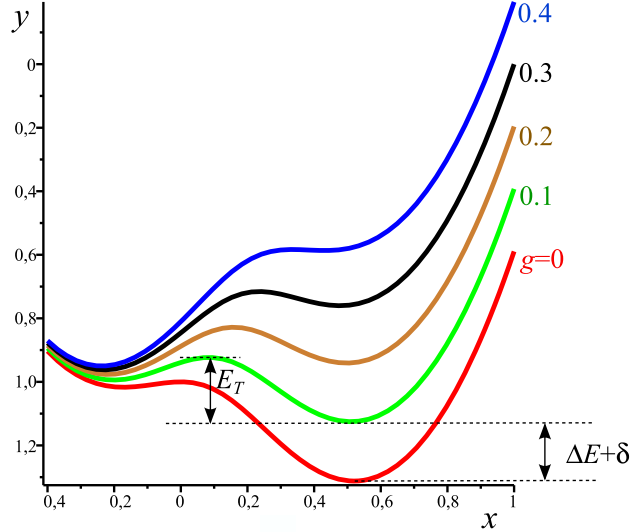

**Fig. 2**. Cross section along the $Q_3$ axis of the $^3T_{1u,z}^*$ potential surface as given by Eq. (1) evaluated for $A=3.5$ and $g$ comprised between zero (in this case we have the cross section of the $^3A_{1u}$ trap level) and $g=0.4$. The barrier height $E_T$ is indicated for $g=0.1$, while the energy separation $\Delta E+\delta$ is indicated between minima relative to the $g=0.1$ and $g=0$ levels. Both quantities $E_T$ and $\Delta E++\delta$ are normalized to $\zeta$.

In order to analyze the characteristic of the delay time in emission, we have to consider the probability of crossing the barrier in two opposite limits: very low and relatively high temperatures. In the first case, the main contribution comes from the tunneling process.

Under the assumption that the potential profile in the region of interest can be considered as cubic, and within the $\theta \to 0$ limit of temperature, the probability of tunneling is given by [9]

$$\Gamma_{\text{tun}}^{-1} = \Omega \left(\frac{30 S_B}{\pi \hbar}\right)^{1/2} \exp\left(-\frac{S_B}{\hbar}\right), \qquad (2)$$

where the action integral $S_B/\hbar = 7.2 \, E_T/\hbar\Omega$, and $\Omega=2\pi\nu$ is the angular frequency of small oscillation inside the potential well around $T_z^*$. The corresponding time delay is given by $\tau_d = \Gamma^{-1}$.

Values of $E_T$, $\Delta E + \delta$ and $\Gamma_{\text{tun}}^{-1}$, multiplied by $\zeta=0.69$ eV, are reported in Tab. 1, for $g = 0.15$-0.25 and $A = 3.5$, and for $g = 0.05$-0.15 and $A = 4$. It is evident that even a small variation in the $A$ parameter causes strong variations in the barrier height and level separation.

The probability of escape from the potential well can also be evaluated by means of the simple WKB relation

$$\Gamma_{\text{WKB}} = \nu \, \exp\left(-\frac{2\pi E_T}{\hbar\Omega}\right), \qquad (3)$$

which supplies results of the same order of magnitude as those of Eq. (2).

With increasing temperature we can define the inversion temperature $\theta_i$ at which the probability of thermal overcoming of the barrier becomes equal to the inverse of the relative lifetime $\tau_{rad}$ from the $T_z^*$ minimum, according to the WKB relation[1]

$$\tau_{rad}^{-1} = \nu \exp\left(-\frac{E_T}{k\theta_i}\right). \tag{4}$$

where $k$ is the Boltzmann constant.

**Table 1**. Determination of the barrier height $E_T$, and level separation $\Delta E + \delta$ for different values of parameters $A$ and $g$. The barrier height and the level separation are evaluated by Eq. (1) and by the inversion temperature $\theta_i$ as given by Eq. (3), which, for $\tau_{rad}$ = 2.5×10$^{-8}$ s and $\nu$ = 4.3×10$^{12}$ s$^{-1}$, is found to be almost coincident with $E_T$ expressed in meV. In the last column, the tunneling probability $\Gamma_{tun}^{-1}$ is given for $\theta\to 0$.

| $A$ | $g$ | $E_T \simeq \theta_i$ (meV) (K) | $\Delta E + \delta$ (meV) | $\Gamma_{tun}^{-1}(\theta \to 0)$ (s) |
|---|---|---|---|---|
| 3.5 | 0.15 | 106 | 189 | 4.78×10$^3$ |
| 3.5 | 0.2 | 76 | 258 | 34.7×10$^{-3}$ |
| 3.5 | 0.25 | 53 | 318 | 4.32×10$^{-6}$ |
| 4 | 0.05 | 90 | 64 | 8.61 |
| 4 | 0.1 | 64 | 128 | 0.31×10$^{-3}$ |
| 4 | 0.15 | 42 | 193 | 5.78×10$^{-8}$ |

In looking at Tab. 1, we realize that the most suitable situation in the case of $A$=3.5, for $\theta_i$ around 80 K, and delay time $\Gamma_{tun}^{-1}$ of the order of milliseconds, is obtained for $g$=0.2. However, the resulting level separation $\Delta E = 241$ meV (for $\delta = 17$ meV [5]) is too far from the value of $\sim 40$ meV as resulting from the experiments of [11,12]. A more favorable situation is obtained with $A$=4 and $g$=0.1 since, still for acceptable values of $\Gamma_{tun}^{-1}$ and $\theta_i$, the level separation $\Delta E$ lowers to 111 meV; however, only for $g = 0.05$, the exact value of $\Delta E = 47$ meV is almost obtained. The resulting high value of $\Gamma_{tun}^{-1} = 8.61$ s is not discouraging, since even a small increase in the temperature immediately lowers the resulting time up to values of the order of milliseconds.

---

[1] A more accurate determination of the non-radiative transition probability can be performed by considering the thermal average over all the involved vibrational levels inside the initial well [10].